\documentstyle[twoside,fleqn,npb]{article}
\newcommand{\eps}{\epsilon}

\newcommand{\AmS}{{\protect\the\textfont2
  A\kern-.1667em\lower.5ex\hbox{M}\kern-.125emS}}

\title{A numerical method for NNLO calculations}

\author{Gudrun Heinrich\address{IPPP, Department of Physics, 
University of Durham, Durham DH1 3LE, England}\thanks{Talk presented at 
RADCOR/Loops \& Legs 2002, Kloster Banz, Germany, September 2002.}
}

\begin{document}

\begin{abstract}
A method to isolate the poles of dimensionally regulated multi-loop integrals 
and to calculate the pole coefficients numerically  
is extended to be applicable to phase space integrals as well. 
\end{abstract}

\maketitle

\section{Introduction}
In the last decades, the interplay between increasing precision from the experimental side
and NLO predictions becoming available from the theoretical side has led to impressive
tests of the Standard Model. 
However, 
certain processes
or observables in QCD should be known at NNLO in perturbation theory in order to match 
the experimental precision, and in the domain of electroweak interactions, many NLO 
processes still await for being calculated exactly.

The calculation of radiative corrections for electroweak processes is mainly complicated
by the presence of several mass scales, whereas in QCD the challenge comes from the
presence of infrared singularities due to massless partons. If
we have to deal with a combination of strong and electroweak interactions, i.e. 
with singularities {\it and} multi-scale problems, the situation is even more involved, 
and the 
complexity of the corresponding analytic calculations 
is enormous. 
On the other hand, the performance of computers is improving continuously, and in order to
calculate cross sections one generally has to use numerical integration at some point 
in any case. 
All this points to the fact that a numerical evaluation of the ingredients 
needed to calculate cross sections beyond leading order, like loop integrals 
and certain phase space integrals, will be of increasing importance in the future. 

In this article I will first present a  method which has been developed
in~\cite{secdec} to evaluate IR (or UV) divergent multi-loop integrals by extracting the
poles in $1/\epsilon$ and calculating the pole coefficients numerically. 
Then I will outline an extension of this method which makes it applicable to the calculation 
of phase space integrals as well.  
Examples from $e^+e^-\to$ 2 jets will illustrate the procedure.

\section{The method}
The method, called "sector decomposition"~\cite{hepp}, is a systematic way 
to disentangle overlapping divergent regions in parameter space. 
The algorithm which has been developed to automate this procedure consists of four 
basic building blocks which will be outlined in the following. 

Consider a scalar graph $G$ with  $N$ propagators and $L$ 
$D$-dimensional loop momenta, typically a master integral. 
The propagators can have powers $\nu_j > 1$
and not necessarily integer. After Feynman parametrisation, 
the graph can be written as 
 \begin{eqnarray}\label{EQ_mixed_rep}
G &=&  
\frac{\Gamma(N_{\nu})}{\prod_{j=1}^{N}\Gamma(\nu_j)}
\int \prod\limits_{j=1}^{N}\,dx_j \,\,x_j^{\nu_j-1}\, \delta(1-\sum_{j=1}^N
x_j)\nonumber\\
&& d{\cal K}_1\dots d{\cal K}_L \left[ 
       \sum\limits_{j,l=1}^{L} k_j \, M_{jl} k_l- 
       2\sum\limits_{j=1}^{L} k_j\, Q_j +J 
                                        \right]^{-N_{\nu}}\nonumber\\
&&\nonumber\\
&=&\frac{(-1)^{N_{\nu}}}{\prod_{j=1}^{N}\Gamma(\nu_j)}
\Gamma(N_{\nu}-LD/2)\int
\limits_{0}^{\infty} \prod\limits_{j=1}^{N}\,dx_j\,x_j^{\nu_j-1} 
\nonumber\\
&&\delta(1-\sum_{l=1}^N x_l)\,\frac{{\cal U}^{N_{\nu}-(L+1) D/2}}
{{\cal F}^{N_{\nu}-L D/2}}\label{eq1}\\
 \nonumber\\
N_{\nu}&=&\sum_{j=1}^N\nu_j\quad ,\quad
d{\cal K}_j=	\frac{d^Dk_j}{i\pi^{\frac{D}{2}}}\nonumber\\		
{\cal F}(\vec x) &=& \det (M) 
\left[ \sum\limits_{j,l=1}^{L} Q_j\cdot Q_l \, M^{-1}_{jl}-J
 \right]\nonumber\\
{\cal U}(\vec x) &=& \det (M) \;.\nonumber	
\end{eqnarray} 
A necessary condition
for the presence of infrared divergences is ${\cal F}=0$, whereas ${\cal U}=0$
can only lead to an UV divergence. 
Upon integration over Feynman parameters, 
both IR and UV poles will manifest themselves as poles in 
1/$\epsilon^a$ ($a\leq 2L$), stemming from endpoint singularities ($x_i=0$) 
of the parameters. 
However, the singular regions are in general  overlapping, 
such that a local subtraction procedure for the poles cannot be immediately applied. 
Our algorithm iteratively separates the overlapping regions in parameter space by the 
following steps:
\subsection*{Part I \quad Generation of primary sectors}
The integration domain is split into
$N$ "primary sectors" by introducing $\theta$-functions:
\begin{eqnarray*}
\int_0^{\infty}d^N x &=& \sum\limits_{l=1}^{N} \int_0^{\infty}d^N x
\prod\limits_{\stackrel{j=1}{j\ne l}}^{N}\theta(x_l- x_j)\theta(x_l)\\
&\Rightarrow & G=\sum_{l=1}^N G_l\;.
\end{eqnarray*}   
Then  the $\delta$--distribution in Eq.~(\ref{eq1}) is eliminated in such a way that the remaining 
integrations are from 0 to 1, which can be achieved by the substitution 
$x_j=x_l t_j$ for $j<l$, $x_j=x_l$ for $j=l$ and $x_j=x_l t_{j-1}$ for $j>l$. 
Because of general homogeneity properties of 
${\cal F}$ and ${\cal U}$, $x_l$ factorises like 
\begin{equation}
{\cal U}(\vec x) \rightarrow {\cal U}_l(\vec t\,)\, x_l^L \quad , \quad
{\cal F}(\vec x) \rightarrow {\cal F}_l(\vec t\,)\, x_l^{L+1}\label{scale}
\end{equation}
and thus, using $\int \frac{dx_l}{x_l}\,\delta(1-x_l(1+\sum_{k=1}^{N-1}t_k ))=1$, one obtains 
in each primary sector $l$
\begin{eqnarray*}
 G_l &=& \int\limits_{0}^{1} \prod\limits_{j=1}^{N-1} dt_j 
\frac{ {\cal U}_l^{N_{\nu}-(L+1)D/2}}{ {\cal F}_l^{N_{\nu}-L D/2}} \quad , \quad
l=1,\dots N\;. 
\end{eqnarray*} 
Note that the feature that singularities only occur as some Feynman parameters
go to zero is preserved by the transformations above. 

\subsection*{Part II \quad Iterated sector decomposition} 
As after part I the overlapping regions in general are not disentangled yet, 
i.e. ${\cal F}$ or ${\cal U}$ still vanish if a certain set of parameters goes 
to zero, the decomposition into sectors is iterated until 
${\cal F}_l$ and ${\cal U}_l$ 
contain a constant term. This iteration produces $k$ new subsectors of
a given primary sector $l$, and in each such subsector one has integrals of the form 
\begin{eqnarray}
G_{lk} &=& \int\limits_{0}^{1} 
\left( \prod\limits_{j=1}^{N-1} dt_j\,\,t_j^{A_j-B_j\epsilon}  \right)
\frac{{\cal U}_{lk}^{N_{\nu}-(L+1)D/2}}{{\cal F}_{lk}^{N_{\nu}-LD/2}}\;,\nonumber\\
&&\label{EQ:subsec_form}
\end{eqnarray}
where now all singularities are factored out in the bracket, while ${\cal F}_{lk}$ and
${\cal U}_{lk}$ do not lead to singularities anymore.
 
\subsection*{Part III \quad Extraction of the poles}
$A_j<0$ in (\ref{EQ:subsec_form}) leads to poles which have to be subtracted. 
Typical for gauge theories are logarithmic singularities, i.e. $A_j=-1$, where 
the subtractions are of the form 
\begin{eqnarray*}
&&\int\limits_0^1 dt_j\, t_j^{-1+B_j\epsilon}\, {\cal I}(t_j,\epsilon)=\hspace*{6cm}\\
&& \frac{{\cal I}(t_j=0,\epsilon)}{B_j\epsilon} 
 +\int\limits_{0}^{1} dt_j \, t_j^{B_j
\epsilon}\,\,\frac{{\cal I}(t_j,\epsilon)-{\cal I}(0,\epsilon)}{t_j} \;.
\end{eqnarray*}
For $A_j<-1$, the Taylor expansion around $t_j=0$ has to be carried out to 
a higher order, but the procedure works analogously. \\
After having isolated the poles in this way, the resulting expression can be 
expanded in $\epsilon$. This leads to a Laurent series for each subsector
integral $G_{lk}$, where the expansion in $\epsilon$ can in principle be carried out 
to arbitrary order $m$: 
$$G_{lk} = \sum\limits_{j=-m}^{2L} C_{lk,j}\,\epsilon^{-j}\;.$$ 
Finally, the $R_l$ subsectors and the $N$ primary sectors are summed over 
to obtain the result for the graph $G$:
$$G=\sum\limits_{l=1}^N G_l=\sum\limits_{l=1}^N\sum\limits_{k=1}^{R_l} G_{lk}\,.$$

\subsection*{Part IV \quad Calculation of the pole coefficients}
What remains to be done is the calculation of the pole coefficients 
$C_j=\sum\limits_{l=1}^N\sum\limits_{k=1}^{R_l} C_{lk,j}\,.$ 
These are ($N-1-\tilde{\j}$)--dimensional  parameter integrals 
($\tilde{\j}$=max(j,0)). In principle, one 
can attempt to perform these integrations analytically, 
and doing this in an automatized way poses no problem 
for large values of $j$, i.e. the leading and subleading poles. 
However, for smaller values of $j$, automatic analytic integration with standard 
algebraic integration routines is not possible anymore. 
On the other hand, numerical integration of these functions does not pose any 
problem as long as the function ${\cal F}$ has no singularities\footnote{In contrast
to the $1/\eps$ poles which have been extracted already, these are 
integrable singularities, like for example threshold singularities, but they 
lead to peaks in multi-parameter space which may cause numerical problems.} 
within the integration region. 
Therefore, for integrals with more than one scale, the  numerical points to be 
calculated are chosen to be 
in the Euclidean region in order to insure that ${\cal F}$ is a regular function. 
In order to deal with general physical kinematics, more sophisticated 
numerical integration methods have to be used, as for example the ones 
suggested in \cite{passarino,thresholds}, 
which can deal with thresholds in the multi-leg one-loop case. 

\section{Examples for loop integrals}
The method outlined above has been applied to check the results for the 
planar\,\cite{Smirnov:1999gc}
and non-planar\,\cite{Tausk:1999vh} on-shell massless double box. 
A prediction\,\cite{secdec}
for the master integrals of the massless double box with one external leg off-shell 
has also been made, which has been confirmed by analytical calculations
later~\cite{Smirnov:2000vy,Gehrmann:2000zt,Smirnov:2000ie,Gehrmann:2001ck}. 
Massless double boxes with two 
off-shell external legs also have been calculated. For example, for the planar
double box with $p_3$ and $p_4$ off-shell, one obtains analytically for the leading and
subleading poles:
\begin{eqnarray*}
&&DB_{m_3m_4} = \Gamma^2(1+\epsilon)\,\frac{(-m_4^2)^{-2\epsilon}}{s^2 t}
\Big\{\;\frac{1}{4\epsilon^4} - \frac{1}{\epsilon^3}\times\\
&&
\Big[\,\frac{1}{2}\log(s/m_4^2)+\log(t/m_4^2)-\frac{1}{2}\log(m_3^2/m_4^2)
\Big] \\
&&+ \quad{\cal O}(\frac{1}{\epsilon^2})\,\Big\}\quad,\quad (m_{3,4}^2=p_{3,4}^2)\;.
\end{eqnarray*}
Numerically e.g. at the point\\
$(-s,-t,-u,-m_3^2,-m_4^2)=(2/3,2/3,2/3,1,1)$, the result is
\begin{eqnarray*}
DB_{m_3m_4}(2/3,2/3,2/3,1,1) = \Gamma^2(1+\epsilon)\qquad\qquad\\
\Big(
                              -\frac{0.8437}{\epsilon^4} 
                              -\frac{2.0524}{\epsilon^3} 
                              -\frac{10.52}{\epsilon^2} 
                              -\frac{48.62}{\epsilon} 
                              - 140.77\Big)\;.
\end{eqnarray*} 
The accuracy is better than 3\% and can be improved easily at the expense of more 
integration time. 

The most challenging loop integral tackled with this method is the planar 
3-loop box. It has been calculated recently 
by Smirnov~\cite{Smirnov:2002mg} who achieved a fully analytical result
for the coefficients of $1/\eps^j\,, j=2\ldots 6$. The numerical result
obtained with our method is in agreement with~\cite{Smirnov:2002mg}.

\section{General multi-parameter integrals}

As the method of sector decomposition is very general and straightforward, 
it can be applied to integrals other than loop integrals, as for 
example phase space integrals, as well. 
However, some properties which are specific to loop integrals, and 
thus were built into the code for loop integrals, 
cannot be used anymore in the case of general parameter integrals:
{\it i)} there is not necessarily a $\delta(1-\sum_{l=1}^N x_l)$ constraint as in
(\ref{eq1}), {\it ii)} the universal scaling properties (\ref{scale}) for ${\cal F}$ and 
${\cal U}$ are lost, and -- most importantly -- {\it iii)} the singularities in $\epsilon$
may not only be at $x_i=0$ anymore. Taking into account {\it i)} and {\it ii)} is not 
a big issue, although it lengthens the code substantially. 
On the other hand, the absence of singularities other than the ones at $x_i=0$
is crucial for the program to work. However, endpoint singularities can always be
remapped by simple variable transformations such that they occur at $x_i=0$ only.
As in the case of loop integrals, singularities {\it within} the integration region have to be 
avoided, for instance by
transforming the integrand correspondingly before feeding it into the code. 

\section{Example from $e^+e^-\to $ 2 jets at NNLO}

In order to calculate a cross section like $e^+e^-\to $ 2 or 3 jets at NNLO, one needs the
following ingredients: The two-loop virtual part, 
single  radiation from
one-loop graphs, and double radiation from tree graphs. 
A lot of progress has been achieved in the past 2 years concerning the first two items. 
For the case of double radiation where two particles are unresolved, 
the soft and collinear limits are known, but 
a systematic procedure to set up a local subtraction scheme which allows 
to isolate and 
analytically integrate the infrared singular regions in phase space has not been
established yet. 
Here it is shown that our code is able to isolate the poles 
and integrate the pole coefficients numerically.
The following IR singular limits can be distinguished:\\
1. three particles collinear\\
2. two pairs of particles collinear\\
3. two particles collinear, one soft\\
4. two particles soft\\
5. a soft $q\bar{q}$ pair. \\
As a specific example, we will consider the triple collinear limit. 
In this limit, phase space and matrix element factorise in the following way:
\begin{eqnarray*}
d\Phi(p_1,p_2,p_3,p_4)&\to & d\Phi^{(2)}(\tilde{p}_{123},\tilde{p}_4)\times d\Phi^{\rm{tc}}\\
|M|^2&\to & |\tilde{M}_B|^2\times \langle P_{g_1g_2q_3}\rangle/s_{123}^2\;,
\end{eqnarray*} 
where $\tilde{p}_{123}$ and $\tilde{p}_4$ are defined in analogy to the 
dipole method~\cite{dipole} for NLO calculations as\\
$\tilde{p}_{123}=p_1+p_2+p_3-\frac{y}{1-y}\,p_4\,,\,
\tilde{p}_4=\frac{1}{1-y}\,p_4\,,\,
y=s_{123}/\tilde{s}\quad (\tilde{s}=(p_1+p_2+p_3+p_4)^2=(\tilde{p}_{123}+\tilde{p}_4)^2)$,
$\tilde{M}_B$ is the `Born' $1\to 2$ matrix element 
with final state momenta $\tilde{p}_{123}$, $\tilde{p}_4$ 
and $\langle P_{g_1g_2q_3}\rangle$ is the spin-averaged triple collinear 
$q\to g_1g_2q_3$ splitting function. 
The triple collinear phase space factor can be cast into the following form ($d=4-2\eps$)
\begin{eqnarray}
&&\int
d\Phi^{\rm{tc}}=\frac{\tilde{s}^{2-2\eps}}{(4\pi)^d}\frac{1}{\Gamma^2(1-\eps)}\qquad\nonumber\\
&&\int dz_1\,dz_2\,dz_3\,\delta(1-\sum_{i=1}^3 z_i)\,[z_1z_2z_3]^{-\eps}\nonumber\\
&&\int_0^1 dy\,[y(1-y)]^{1-2\eps}\int_0^1 du\,[u(1-u)]^{-\eps}\nonumber\\
&&\frac{\Gamma(1-\eps)}{\sqrt{\pi}\Gamma(\frac{1}{2}-\eps)}
\int_0^{\pi}d\theta (\sin\theta)^{-2\eps}\label{phitc}
\end{eqnarray}
where
$z_i=p_i\tilde{p}_4/\tilde{p}_{123}\tilde{p}_4\quad (i=1,2,3)$
and the integral over $u$ is a rescaled transverse momentum integration. 
In the above parametrisation, the Mandelstam variables are given by:
\begin{eqnarray}
s_{123}&=&\tilde{s} y\nonumber\\
s_{13}&=&\tilde{s} y u (1-z_2)\nonumber\\
s_{23}&=&\tilde{s} y u
\frac{z_1z_2}{(1-z_2)}(1+\nu^2-2\nu\cos{\theta})\nonumber\\
s_{12}&=&\tilde{s} y u
\frac{z_2z_3}{(1-z_2)}(1+\lambda^2+2\lambda\cos{\theta})\nonumber\\
s_{i4}&=&\tilde{s} (1-y) z_i\quad (i=1,2,3)\nonumber\\
\nu^2&=&\frac{z_3}{z_1z_2}\frac{(1-u)}{u}
\quad,\quad \lambda^2=\frac{z_1^2}{z_3^2}\,\nu^2\label{mandel}
\end{eqnarray}
The triple collinear splitting functions can be found in~\cite{camp,Catani:1999ss}. 
The aim here is only to demonstrate how the method operates. So let us focus
first on the Abelian part of the splitting  $q\to g_1g_2q_3$, described
by the splitting function 
\begin{eqnarray}
&&\langle P^{Abel}_{g_1g_2q_3}\rangle/s_{123}^2=C_F^2\,\Biggl\{
\frac{A(z)}{s_{13}s_{23}}\nonumber\\
&&+\frac{B_1(z)}{s_{13}s_{123}}+
\frac{B_2(z)}{s_{23}s_{123}}\nonumber\\
&&
+C\,\left(\frac{s_{23}}{s_{13}s_{123}^2}+
\frac{s_{13}}{s_{23}s_{123}^2}\right)+\frac{D}{s_{123}^2}\Biggr\}\;.\nonumber
\end{eqnarray}
We want to integrate this function over the triple collinear phase space factor given
in (\ref{phitc}). Using parametrisation (\ref{mandel}) for the Mandelstam variables, 
we see that the $y$-integration factorises immediately, and that the 
$u$-integration also factorises except if there is an angular dependence through 
$s_{23}$. In this case, however, one has to insure that no (integrable) 
singularity in the analytic plane is crossed before 
feeding the function into the numerical integration routine. 
This can be achieved easily by doing the $\theta$-integration analytically. 
Using 
\begin{eqnarray}
&&\frac{\Gamma(1-\eps)}{\sqrt{\pi}\Gamma(\frac{1}{2}-\eps)}
\int_0^{\pi}d\theta (\sin\theta)^{-2\eps}[1+\nu^2\pm 2\nu\cos{\theta}]^{-1}\nonumber\\
&=&\theta(1-\nu^2)\, _2F_1(1,1+\eps,1-\eps,\nu^2)\nonumber\\
&+&
\theta(\nu^2-1)\frac{1}{\nu^2}\,_2F_1(1,1+\eps,1-\eps,1/\nu^2)\label{ia}
\end{eqnarray}
one can split the subsequent integrations into two parts at $\nu^2=1$. 
After remapping to integrals from 0 to 1 and using the integral representation for
the Hypergeometric function, one arrives at the `standard' form required by the
numerical integration routine. 
The result is given by
\begin{eqnarray}
I^{Abel}&=&\int d\Phi^{\rm{tc}}\langle P^{Abel}_{g_1g_2q_3}\rangle/s_{123}^2\nonumber\\
&=&\frac{\tilde{s}^{-2\eps}}{(4\pi)^d}\frac{C_F^2}{\Gamma^2(1-\eps)}\sum_{i=1}^4\frac{P_i}{\eps^i}\nonumber\\
&&\nonumber\\
P_4 &=& 4	\pm 2\times 10^{-6}\nonumber\\
P_3 &=& 13.99999	\pm 4.8\times 10^{-5}\nonumber\\
P_2 &=& 10.1125    \pm 0.0015\nonumber\\
P_1 &=& -38.8109     \pm 0.012\nonumber\\
P_0 &=&  -116.47    \pm 0.07\nonumber
\end{eqnarray}
The same procedure can be followed for the other singular limits of the matrix
element. 

The most complicated denominator occurs in the non-Abelian part of the triple 
collinear splitting function, where an angular dependent denominator 
appears quadratically, \\
$ \langle P^{\mathrm{non-Ab}}_{g_1g_2q_3}\rangle/s_{123}^2\sim
A^{\prime}(z)/s_{12}^2$ + less complicated. 
However, this does not present a problem for the numerical method. 
Again, all one has to do is to insure that no singularities in the analytic plane 
are crossed. Carrying out the angular integration leads to
\begin{eqnarray}
I_{\theta}&=&\frac{\Gamma(1-\eps)}{\sqrt{\pi}\Gamma(\frac{1}{2}-\eps)}
\int_0^{\pi}d\theta (\sin\theta)^{-2\eps}\nonumber\\
&&[1+\lambda^2+ 2\lambda\cos{\theta}]^{-2}
=\nonumber\\
&&\theta(1-\lambda^2)\,\, _2F_1(2,2+\eps,1-\eps,\lambda^2)\nonumber\\
&+&
\theta(\lambda^2-1)\frac{1}{\lambda^4}\,\,_2F_1(2,2+\eps,1-\eps,1/\lambda^2)\nonumber
\end{eqnarray}
Following the same procedure as outlined for (\ref{ia}) again leads to a 
representation which can be fed into the numerical routine.

\section{Conclusions}

A constructive algorithm to isolate infrared singularities from 
multi-loop integrals and phase space integrals has been presented. The algorithm
produces finite parameter integrals as pole coefficients, which can be integrated 
numerically, or analytically in simple cases. For the numerical integration, 
one has to insure that the function  does not have (integrable) 
singularities within the integration region, 
which can be achieved easily by appropriate variable transformations in the case of 
phase space integrals with one overall scale.
As applications, a result 
for a massless double box with two legs off-shell and an example from the 
phase space of $e^+e^-\to 2$ jets have been given. 
The method can serve to check various kinds of analytical NNLO results numerically
and also provides a step towards a completely numerical evaluation of 
radiative corrections.

\medskip

{\bf Acknowledgements}\\
I would like to thank V.~A.~Smirnov for the exchange of results which 
triggered further developments of the program, and the organisers
of the conference for the kind invitation.


\begin{thebibliography}{9}

\bibitem{secdec} T.~Binoth and G.~Heinrich,
Nucl.\ Phys.\ B {\bf 585} (2000) 741.

\bibitem{hepp}K.~Hepp, 
Commun.\ Math.\ Phys.\  {\bf 2}, 301 (1966);\,
see also contribution of S.~Pozzorini in these proceedings.

\bibitem{passarino}
G.~Passarino,
Nucl.\ Phys.\ B {\bf 619} (2001) 257;\\
A.~Ferroglia, M.~Passera, G.~Passarino and S.~Uccirati,
hep-ph/0209219.

\bibitem{thresholds}
T.~Binoth, G.~Heinrich and N.~Kauer,
hep-ph/0210023.

\bibitem{Smirnov:1999gc}
V.~A.~Smirnov,
Phys.\ Lett.\ B {\bf 460} (1999) 397.
\bibitem{Tausk:1999vh}
J.~B.~Tausk,
Phys.\ Lett.\ B {\bf 469} (1999) 225.

\bibitem{Smirnov:2000vy}
V.~A.~Smirnov,
Phys.\ Lett.\ B {\bf 491} (2000) 130.
\bibitem{Gehrmann:2000zt}
T.~Gehrmann and E.~Remiddi,
Nucl.\ Phys.\ B {\bf 601} (2001) 248.

\bibitem{Smirnov:2000ie}
V.~A.~Smirnov,
Phys.\ Lett.\ B {\bf 500} (2001) 330.
\bibitem{Gehrmann:2001ck}
T.~Gehrmann and E.~Remiddi,
Nucl.\ Phys.\ B {\bf 601} (2001) 287


\bibitem{Smirnov:2002mg}
V.~A.~Smirnov, in these proceedings [hep-ph/020929];
see also Phys.\ Lett.\ B {\bf 547} (2002) 239. 

\bibitem{dipole}S.~Catani and M.~H.~Seymour,
Nucl.\ Phys.\ B {\bf 485} (1997) 291
[Erratum-ibid.\ B {\bf 510} (1997) 503].

\bibitem{camp}
J.~M.~Campbell and E.~W.~Glover,
Nucl.\ Phys.\ B {\bf 527} (1998) 264.
\bibitem{Catani:1999ss}
S.~Catani and M.~Grazzini,
Phys.\ Lett.\ B {\bf 446} (1999) 143;
S.~Catani and M.~Grazzini,
Nucl.\ Phys.\ B {\bf 570} (2000) 287.



\end{thebibliography}
\end{document}